\documentclass[aps,prb,twocolumn,amsmath,amssymb,superscriptaddress,floatfix]{revtex4}
\usepackage{graphicx}
\bibstyle{apsrev.bib}

\usepackage{subfigure}

\newcommand{\bc}{\begin{center}}
\newcommand{\ec}{\end{center}}
\newcommand{\be}{\begin{equation}}
\newcommand{\ee}{\end{equation}}
\newcommand{\ba}{\begin{array}}
\newcommand{\ea}{\end{array}}
\newcommand{\beqn}{\begin{eqnarray}}
\newcommand{\eeqn}{\end{eqnarray}}

\begin{document}

\title{Nonequilibrium critical dynamics of the two-dimensional Ising model quenched from a
correlated initial state}

\author{L\'aszl\'o K\"ornyei}
 \affiliation{Institute of Theoretical Physics,
Szeged University, H-6720 Szeged, Hungary}
\author{Michel Pleimling}
\email{Michel.Pleimling@vt.edu}
 \affiliation{Department of Physics, Virginia Polytechnic Institute and State University,
 Blacksburg, Virginia 24061-0435, USA}
\author{Ferenc Igl\'oi}
 \email{igloi@szfki.hu}
 \affiliation{Research Institute for Solid State Physics and Optics,
H-1525 Budapest, P.O.Box 49, Hungary}
 \affiliation{Institute of Theoretical Physics,
Szeged University, H-6720 Szeged, Hungary}

\date{\today}

\begin{abstract}
The universality class, even the order of the transition, of the two-dimensional Ising model
depends on the range and the symmetry of
the interactions (Onsager model, Baxter-Wu model, Turban model, etc.), but the
critical temperature is generally the same due to self-duality. Here we consider
a sudden change in the form of the interaction and study the nonequilibrium
critical dynamical properties of the nearest-neighbor model. The relaxation of
the magnetization and the decay of the autocorrelation function are found to
display a power law behavior with characteristic exponents that depend on the
universality class of the initial state.

\end{abstract}


\maketitle

\section{Introduction}
The parameters of a physical system are often subject to a sudden change, as for example
fast cooling or heating or the switching on or off of an external field. In material science
these processes are used to prepare new types of (glassy) states of matter. After such a
quench the systems are out of equilibrium and their dynamical properties can be completely different
from those known in the equilibrium situation. The phenomena of aging and rejuvenation are
typical examples of such glassy dynamics\cite{kr03,buch1}.

In a theoretical approach a simple and often studied example\cite{godr,ritort,cg04} is a $d$-dimensional
ferromagnet, which is quenched
from its high-temperature (paramagnetic) phase to a temperature $T$ which is below or at
the critical temperature $T_c$. For $T<T_c$ we have phase-ordering kinetics where order
grows through domain coarsening\cite{bray}. For a critical quench, $T=T_c$, the domains are fractals, and
the growth and dynamics involve critical exponents\cite{fisher} such as the
magnetization scaling dimension $x=\beta/\nu$ ($\beta$ and $\nu$ are the critical exponents
of the magnetization and the correlation length, respectively)
and the dynamical exponent $z$.

In this case one often measures the autocorrelation function\cite{huse89},
$G(t,s)=\langle \sigma(t)\sigma(s) \rangle$,
where $\sigma(t)$ is the operator of the magnetization and $t$ and $s$ are the observation
time and the waiting time, respectively. Aging is manifested by the fact that $G(t,s)$ is
nonstationary, but instead has the scaling form $G(t,s)=t^{-2x/z} \tilde{G}(s/t)$, where the scaling
function for small argument behaves as $\lim_{\tau \to 0} \tilde{G}(\tau)\sim \tau^{(d-x_i-x)/z}$.
Here $x_i$ is a new nonequilibrium exponent, the anomalous dimension of the initial magnetization\cite{jans92}.
The autocorrelation function for $t \gg s$ behaves as $G(t) \sim t^{-\lambda/z}$, where
the autocorrelation exponent satisfies the relation $\lambda=d-x_i+x$.
Another common measurement after a critical quench concerns the relaxation of the magnetization\cite{jss89},
$m(t)=\langle \sigma(t) \rangle$, where the system is initially prepared in a state with a small initial
magnetization $m_i$. Here we have asymptotically $m(t) \sim t^{\theta}$, where the
initial slip exponent is given by $\theta=(x_i-x)/z$.

In critical nonequilibrium dynamics the initial state might contain some kind of correlations.
In the case of a perfectly correlated initial state the relaxation process involves only
equilibrium exponents. Here we have $m(t) \sim t^{-x/z}$, and, similarly, for
$t \gg s$, $G(t) \sim t^{-x/z}$. The dynamic crossover between the ordered and the 
disordered cases has been the subject of a recent series of papers \cite{Cal06,Cal07,Pau07}.
Another possibility is given by initial states that display
quasi-long-range order, i.e. where correlations decay as a power law. This happens for the 
two-dimensional XY model
if both the initial temperature, $T_i$, and the final temperature of the quench, $T$, are below the 
Kosterlitz-Thouless temperature, $T_{KT}$. If $T_i<T \ll T_{KT}$, we have
according to spin-wave theory\cite{Berthier,abriet}
$G(t) \sim t^{-(x(T)-x(T_i))/z}$, where $x(T)$ is the value of the anomalous dimension at the given
temperature. Other recent studes have focused on the $d$-dimensional spherical model with an initial state
of prescribed correlations\cite{picone} or on the Ising model with initial states generated through random
field effects\cite{epl}.

In the present paper we consider quenches during which the temperature of the
system remains the same, but where the form and the local symmetry of the interactions are changed.
Interestingly, the recent progress in the experiments on phase transitions in optical lattices
could make this type of investigations possible.
We here consider the case of Ising spins on a square lattice
with different types of interactions. 
The phase transition encounterd in the Ising model is considered to 
be the paradigm of an order-disorder transition since the exact solution of Onsager\cite{onsager}
of the standard model with nearest neighbor couplings.
The same model, however, with three-spin product interaction for each elementary triangle
belongs to a different universality class. According to the exact treatment by Baxter and Wu\cite{BW}
the (static) critical exponents of this so called Baxter-Wu (BW) model are the same as those of the
four-state Potts model\cite{baxter82}. Still another self-dual model 
has been introduced by Turban\cite{turban1,n3t,turban}
and others\cite{gruber,n3a}
that has nearest-neighbor interactions in the vertical direction, but $n$-spin product
interactions in the horizontal direction. Of course, for $n=2$ we recover the Onsager problem,
whereas for $n=3$, according to symmetry arguments, approximate mappings and numerical
investigations, the system belongs to the four-state Potts
universality class. For $n \ge 4$ the phase transition is of first order.

In our study we prepare the system in an initial state that is an equilibrium critical
state of the BW-model or of the multispin model with $n=3$ and $n=4$.
In the initial state of the BW-model and that of the $n=3$ model there are
the same type of critical correlations, since the two models belong to the same
universality class. On the other hand for the $n=4$ model we have
phase coexistence at the phase transition point between the ordered and the disordered phases. 
Having prepared the system in this way, we change
at time $t=0$ the form of the interaction and then let the system evolve according
to spin-flip dynamics with nearest neighbor interactions. We thereby measure the relaxation of
the magnetization and the decay of the autocorrelation function and determine the exponents
$\lambda/z$ and $\theta$.

The structure of the paper is the following. In Sec. \ref{sec:model} we define the models
and describe their critical properties. The results of nonequilibrium relaxation are presented
in Secs. \ref{sec:second} and \ref{sec:first} where the initial state corresponds to 
an equilibrium state of a second order
or of a first order transition point, respectively. We discuss our results in the final section.

\section{Models and their critical properties}
\label{sec:model}

We consider in the following Ising spins, $\sigma_{i,j}=\pm 1$, at the sites of a square lattice
with different types of ferromagnetic interactions.

\paragraph{Ising model.}

The standard Ising (or Ising-Lenz) model only contains nearest neighbor ferromagnetic couplings,
so that the Hamiltonian is given by
\be
{\cal H}_I=-\sum_{i,j} J \left( \sigma_{i,j}\sigma_{i,j+1}+\sigma_{i,j}\sigma_{i+1,j} \right)\;.
\label{H_I}
\ee
where $i$ and $j$ label the lattice sites, whereas $J$ is the strength of the couplings.
The critical point is given by the condition\cite{onsager}
\be
\sinh(2J/kT_c)=1\;,
\label{crit_I}
\ee
which separates a two-fold degenerate ordered phase from a paramagnetic phase. The static critical
exponents are known exactly, whereas the dynamical exponent $z$
and the nonequilibrium exponents $\lambda/z$ and $\theta$ are calculated numerically with
high precision. The values of these exponents are collected in Table \ref{table:1}.

\paragraph{Baxter-Wu model.}

In the Baxter-Wu model we have three-spin product interactions with strength $J_{BW}$ between spins located
on elementary triangles, and the Hamiltonian is given by
\be
{\cal H}_{BW}=-\sum_{i,j} J_{BW} \left(
\sigma_{i,j}\sigma_{i-1,j}\sigma_{i,j-1}+\sigma_{i,j}\sigma_{i+1,j}\sigma_{i,j+1}\right)\;.
\label{H_BW}
\ee
The ordered phase of the system is four-fold degenerate,
with the majority spin orientations in the three equivalent sublattices being given by : 
$\uparrow,\uparrow,\uparrow$; $\uparrow,\downarrow,\downarrow$;
$\downarrow,\uparrow,\downarrow$; and $\downarrow,\downarrow,\uparrow$.
According to exact results the critical point of the system is located at\cite{BW}
\be
\sinh(2J_{BW}/kT_c)=1\;,
\label{crit_BW}
\ee
and the static critical exponents are the same as for the four-state Potts model\cite{baxter82},
but without logarithmic corrections to scaling. Numerical results indicate that 
universality also holds for the dynamical exponent, $z$. The nonequilibrium
exponents, $\lambda$ and $\theta$, however, seem to be different, see Table \ref{table:1}.

\paragraph{Turban model.}

In this model we have nearest neighbor interactions with strength $J_2$ in the vertical
direction and $n$-spin product interactions with strength $J_n$ in the horizontal direction, so that the
Hamiltonian is given by
\be
{\cal H}_n=-\sum_{i,j}\left(J_2 \sigma_{i,j} \sigma_{i,j+1} + J_n \prod_{k=0}^{n-1} \sigma_{i+k,j}\right)\;.
\label{H_turban}
\ee
The ordered phase of the system is $2^{n-1}$-fold degenerate.
The model is self-dual \cite{gruber,turban1,turban} and the self-dual point is located at
\be
\sinh(2J_2/k T_c)\sinh(2J_n/k T_c)=1\;.
\label{crit_turban}
\ee
It is known from numerical studies that a single phase transition takes place in the system, thus the
phase transition temperature coincides with the self-duality point. In
the following we take $J_2=J_n$ ($= J = J_{BW}$), so that
all models described in this Section have the same critical temperature, see
Eqs.(\ref{crit_I}), (\ref{crit_BW}) and (\ref{crit_turban}). As already mentioned in
the introduction the Turban model with $n=3$ has a continuous phase transition which
belongs to the (static) universality class of the four state Potts model
\cite{n3t,n3a,n3b,n3c,n3d,n3e}, even the logarithmic corrections are expected to be
of the same form for the two models. The numerical estimates of the dynamical exponent
of the two models are somewhat different, although they could be the same within the
error of the calculation. The same conclusion holds also for the nonequilibrium exponents,
$\lambda$ and $\theta$, see Table \ref{table:1}.

The model for $n \ge 4$ has a first order transition\cite{n4,blote}. Detailed numerical
studies are available for the $n=4$ case which exhibits a latent heat of $\Delta/k T_c=0.146(3)$ and a
jump of the magnetization from zero to $m_c=0.769(6)$. Nonequilibrium relaxation studies of this
model have been performed recently\cite{pi07}. The autocorrelation function
has thereby been found to approach its limiting value, given by
the magnetization in the ordered phase at the transition point, $m_c$, through a stretched
exponential decay. On the other hand relaxation of the
magnetization starting with an uncorrelated initial state with a small magnetization, $m_i$,
has been shown to approach zero with an asymptotic power law time dependence,
thus from a nonequilibrium point of view the transition is continuous.

\begin{table}
\caption{
Upper part: static and dynamic critical quantities of the Ising model, the Baxter-Wu model
and the Turban model with $n=3$ and $n=4$. 
$x$: bulk scaling dimension, $z$: dynamical scaling exponent, $\lambda$: autocorrelation
exponent, $\theta$: initial slip exponent. For $n=4$ the phase transition is of
first order: $^{(a)}$ discontinuity fixed-point value, $^{(b)}$ stretched exponential decay.
Lower part: nonequilibrium critical exponents of the two-dimensional Ising model starting with an initial
state corresponding to the critical state of the Baxter-Wu model
and the Turban model with $n=3$ and $n=4$.
\label{table:1}}
 \begin{tabular}{|c|c|c|c|c|}  \hline
   & $x$ & $z$ & $\lambda/z$ & $\theta$ \\ \hline
  Ising & $1/8$ & $2.17$ & $0.74(2)$\cite{chatelain04} & $0.187$ \\
  BW& $1/8$ & $2.29(1)$\cite{adf03} & $1.13(6)$\cite{chatelain04} & $-0.186(2)$\cite{adf03} \\
  $n=3$ & $1/8$ & $2.3(1)$\cite{sdf01} & $0.98(2)$\cite{chatelain04} & $-0.03(1)$\cite{sdf01} \\
  $n=4$ & $0.^{(a)}$ & $2.05(10)$\cite{pi07} & $\infty^{(b)}$ & $-1.00(5)$\cite{pi07} \\ \hline \hline
  BW&  &  & $0.17(1)$ & $0.18(1)$ \\
  $n=3$ &  &  & $0.165(10)$ & $0.18(1)$ \\
  $n=4$ &  &  & $0.475(10)$ &  \\ \hline

  \end{tabular}
  \end{table}

\section{Relaxation from second order transition points}
\label{sec:second}

In the following we present results of the relaxation of the critical Ising
model where we start from a critical state
of the BW model and of the $n=3$ Turban model. The two models
have the same static critical exponents and, interestingly, the decay of the
critical correlations is the same as in the critical Ising model. We start to
present first the results for the BW model.

\subsection{Relaxation from a BW critical state}
\label{sec:BW}
\begin{figure}[h!]
  \begin{center}
     \includegraphics[width=3.3in,angle=0,clip=]{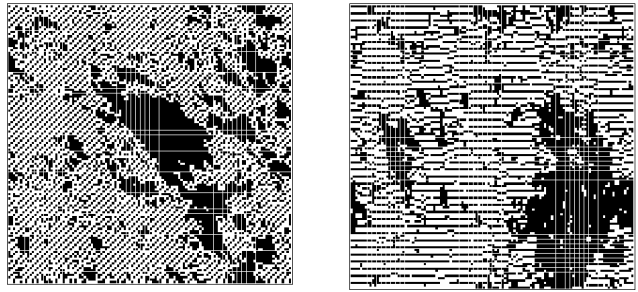}
   \end{center}
   \caption{
Typical initial state of the BW-model (left) and the $n=3$ Turban model (right)
with $L=120$ having a magnetization $m_i=0$.
The initial state is a mixture of the pure phases.}
   \label{Fig:figure1}
 \end{figure}

In the actual calculations we used finite systems composed of $L \times L$ spins, with $L$ 
ranging from 60 to 240.
We set the temperature to the critical value 
given in Eq.(\ref{crit_BW})
and then let the system evolve under the Hamiltonian ${\cal H}_{BW}$, see Eq.(\ref{H_BW}), using
the cluster-flip Monte-Carlo algorithm\cite{sw}. After equilibrium is reached the critical BW-states are
extracted and sorted by magnetizations. For each magnetization, $m_i$, we have selected $1000$ independent
starting states. A typical starting configuration for $m_i=0$ is shown in the left panel of Fig. \ref{Fig:figure1}.

After having selected the initial state, we subject the system to the heat-bath dynamics of the critical Ising
model, i.e. the relaxation is performed at the same temperature as the initialization but the
couplings between the spins are changed. For a given
starting configuration the relaxation is repeated with typically a few hundred independent sets of random numbers.

\begin{figure}[h!]
  \begin{center}
     \includegraphics[width=3.3in,angle=0,clip=]{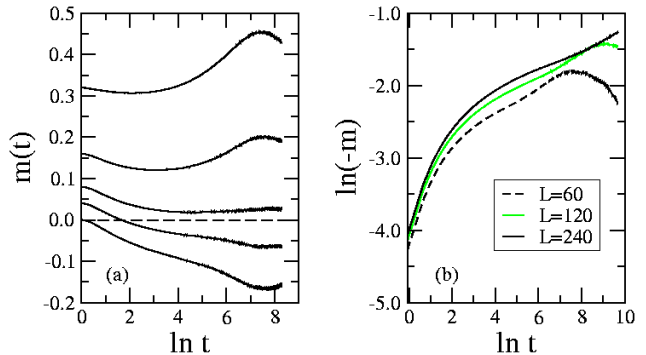}
   \end{center}
   \caption{
(Color online) (a) Relaxation of the magnetization starting with a critical $BW$ state with
different initial magnetizations $m_i$ (from top to bottom, $m_i = 0.33$, 0.16, 0.08, 0.04, and 0). 
The data shown have been obtained for $L=60$.
(b)  Relaxation of the magnetization starting with a critical $BW$ state with
initial magnetization $m_i=0$ in a log-log scale for different system sizes.
}
   \label{Fig:figure2}
 \end{figure}

Fig. \ref{Fig:figure2} shows the time dependence of the magnetization for different
starting magnetizations $m_i$.
It is seen in Fig. \ref{Fig:figure2}a that for small times the magnetization
is monotonously decreasing, yielding for $m_i < 0.08$ negative values. The
asymptotic dependence of $m(t)$ is studied in more details 
in Fig. \ref{Fig:figure2}b for $m_i=0$ where we plot the magnetization
vs. time in a log-log scale. According to this figure the absolute value 
of the magnetization has a power-law dependence. For times shorter than some size-dependent time $t_1(L)$
there is an effective exponent, $\theta_1 =0.13(1)$, which turns to a different value, $\theta=0.18(1)$,
for $t>t_1(L)$.
The available time-scale is restricted by the finite size of the system, through $t_{m} \sim L^z$,
$z$ being the dynamical exponent of the critical Ising model, see Table \ref{table:1}. The cross-over
behavior seen in Fig. \ref{Fig:figure2}b is attributed to the structure of the initial state. As seen
in Fig. \ref{Fig:figure1}a the initial state can be described as a composition of the four different
pure ordered phases of the $BW$ model, one with magnetization, $m=1$, and three others each
with magnetization, $m=-1/3$. In the early time steps these pure phases relax and the relaxation
of the mixture of the $m=-1/3$ phases has an effective exponent, $\theta_1 =0.13(1)$. This
assumption is in accordance with the effective exponent measured in Fig. \ref{Fig:figure2}a for $m_i=1/3$.
Then for $t>t_1(L)$ the remains of the pure $m=-1/3$ phases are dissolved and we are in the true
asymptotic regime. 

\begin{figure}[h!]
  \begin{center}
     \includegraphics[width=3.3in,angle=0,clip=]{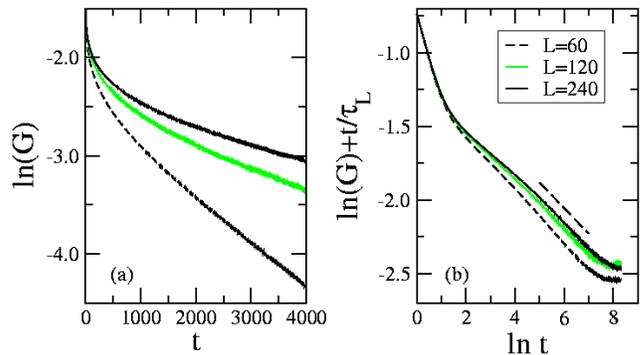}
   \end{center}
   \caption{
(Color online) (a) Autocorrelation function starting from a critical BW initial state for different system
sizes.
In the semi-log plot the asymptotic slope is proportional to the inverse characteristic
time $\tau(L)$. (b) The corrected autocorrelation function, $\tilde{G}(t)=G(t)\exp(t/\tau_L)$,
in a double-logarithmic scale for the same system sizes. 
The slope of the curves corresponds to the exponent $\lambda/z$.
}
   \label{Fig:figure3}
 \end{figure}

Next we consider the nonequilibrium autocorrelation function, $G(t)$, calculated from a
critical BW state with vanishing magnetization, $m_i=0$. The autocorrelation function
is thereby defined by
\be
G(t) = \frac{1}{L^2} \sum\limits_{i=1}^{L^2} \langle \sigma_i(t) \sigma_i(0) \rangle
\ee
where we average both over initial states and different realizations of the noise.
According to the numerical results
shown in Fig. \ref{Fig:figure3}a, the autocorrelation
function in a finite system of linear size $L$ is well described by the functional form
\be 
G(t)=A t^{-\lambda/z} \exp(-t/\tau_L)\;,
\label{G_exp}
\ee
where the characteristic time, $\tau_L$, is a monotonously increasing function of $L$, so
that in the thermodynamic limit $G(t)$ has a power law dependence. We tried to fit
the measured $\tau_L$ in the form $\tau_L \sim L^{\zeta}$ and for small $L$ 
the obtained
exponent $\zeta$ is compatible with the dynamical exponent of the Ising model, as
listed in Table \ref{table:1}. For larger $L$, however, when the characteristic time becomes
larger than $t_1$, the measured
$\zeta$ is found to decrease below $1.5$. A possible explanation of this behavior is that
the relevant length scale in the problem, the typical size of the clusters of pure phases, $\xi$, see
Fig. \ref{Fig:figure1}a is smaller than $L$.

In order to deduce the autocorrelation exponent, $\lambda$, from the numerical data we
have calculated the corrected autocorrelation function, $\tilde{G}(t)=G(t)\exp(t/\tau_L)$, in
which we used the estimate for $\tau_L$ calculated from Fig. \ref{Fig:figure3}a.  $\tilde{G}(t)$ is plotted
in Fig. \ref{Fig:figure3}b in a double-logarithmic scale for different system sizes.
Clearly, the curves are seen to approach an
asymptotic curve for large $L$.

Here, as for the relaxation process, one can observe an
early time period in which case the effective exponent is $\lambda_1/z \approx 0.14(1)$.
In the true asymptotic range the measured autocorrelation exponent is somewhat
larger, $\lambda/z=0.18(1)$. Interestingly, this value is much smaller than the value $0.74$
obtained when starting from a fully disordered initial state.

\subsection{Relaxation from an $n=3$ critical state}

For the case of the $n=3$ Turban model we proceed as for the BW case. 
Using Glauber dynamics, we generate equilibrium critical states of the Hamiltonian
(\ref{H_turban}) with three-spin interactions. 
After equilibrium is reached the initial states are
sorted by magnetization (see Fig. \ref{Fig:figure1}b for an example). These initial states are
then subjected to the heat-bath dynamics of the critical Ising model. The linear sizes of the system
used here are the same as for the BW-relaxation described in Sec. \ref{sec:BW}.

\begin{figure}[h!]
  \begin{center}
     \includegraphics[width=3.3in,angle=0,clip=]{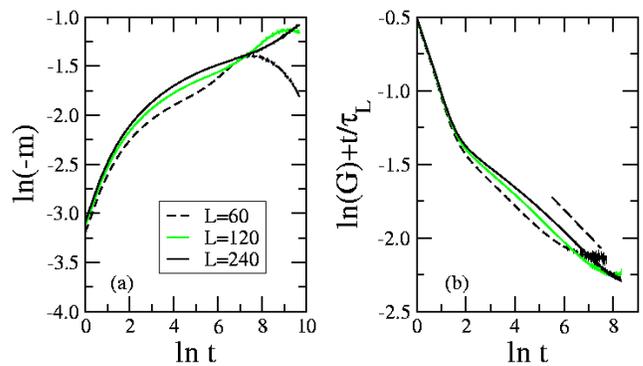}
   \end{center}
   \caption{
(Color online) (a) Relaxation of the magnetization starting with a critical $n=3$ state with
initial magnetization $m_i=0$ in a log-log scale for different system sizes.
(b) The corrected autocorrelation function, $\tilde{G}(t)=G(t)\exp(t/\tau_L)$,
in a double-logarithmic scale for the $n=3$ model. 
The slope of the curves corresponds to the exponent $\lambda/z$.
}
   \label{Fig:figure4}
 \end{figure}

Starting with initial states with $m_i=0$ the magnetization relaxes to negative values.
To obtain a qualitative estimate of the initial slip exponent $\theta$ we plot in Fig. \ref{Fig:figure4}a
the absolute magnetization vs. time in a log-log scale. From the asymptotic slope of this curve we
obtain $\theta=0.18(1)$, which is within error bars the value obtained from a BW initial state 
in Sec. \ref{sec:BW}.

For the autocorrelation function the same ansatz as given in Eq.(\ref{G_exp}) works in this
case too, and the characteristic time, $\tau_L$, has a similar size dependence as for the
BW-model. Having estimated $\tau_L$ for each size we have calculated the corrected
autocorrelation function, $\tilde{G}(t)$, which is plotted in Fig. \ref{Fig:figure4}b in a
log-log scale. From the asymptotic slope of these curves we have estimated the autocorrelation
exponent as $\lambda/z=0.165(10)$, which is consistent , within the error of the calculation, 
with the value obtained from BW initial states in Sec. \ref{sec:BW}.

\section{Relaxation from a first-order transition point}
\label{sec:first}

The Turban model with four-spin product interaction, $n=4$, has an eight-fold degenerate
state in the ordered phase, and this degeneracy is lifted at the first-order transition
point at $T=T_c$. The transition point is 
characterized by the phase coexistence of the paramagnetic phase and
the ordered phase, yielding a state that has a completely different structure as the equilibrium critical states
of the $n=3$ model or the BW model considered in the previous section. Since the order-parameter
is non-zero even at $T=T_c$, we consider in the following only the nonequilibrium autocorrelation function.
Starting with an initial state of the $n=4$ model at $T_c$ we have used heat-bath dynamics
of the critical Ising model.

\begin{figure}[h!]
  \begin{center}
     \includegraphics[width=3.3in,angle=0,clip=]{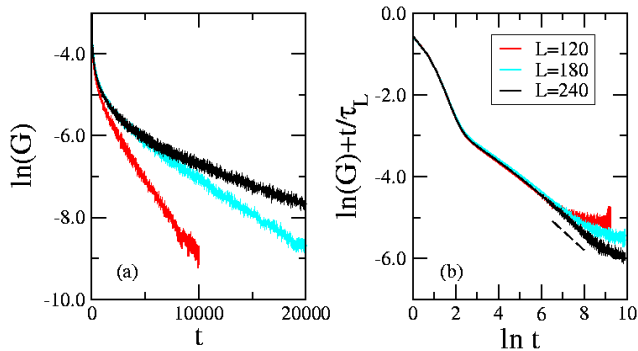}
   \end{center}
   \caption{
(Color online) (a) Autocorrelation function starting from a first-order transition equilibrium state of the
Turban model with four-spin product interaction for different system
sizes.
In the semi-log plot the asymptotic slope is proportional to the inverse characteristic
time, $\tau(L)$. (b) The corrected autocorrelation function, $\tilde{G}(t)=G(t)\exp(t/\tau_L)$,
in a double-logarithmic scale for the same system sizes.
The slope of the curves corresponds to the exponent $\lambda/z$.
}
   \label{Fig:figure5}
 \end{figure}

The nonequilibrium autocorrelation function of this system is shown in Fig. \ref{Fig:figure5}a.
The decay of the autocorrelation is again well described by the same functional form 
(\ref{G_exp}) that we have already observed for the two cases discussed before.
From the system size dependent characteristic times $\tau_L$ we infer a value $z \approx 2.1$ for the
dynamical exponent, in good agreement with the literature value 2.17 for the two-dimensional critical
Ising model. The corrected autocorrelation $\tilde{G}(t)=G(t)\exp(t/\tau_L)$ displays a power-law decay as
shown in Fig. \ref{Fig:figure5}b, but with an exponent $\lambda/z = 0.475(10)$ that is much larger then the value
we have obtained when starting from the critical BW or $n=3$ Turban model. Interestingly, the value of
$\lambda/z$ is compatible with $1/z$, yielding the value $\lambda=1$ for the autocorrelation exponent.

In the following we present a simple scaling argument which could explain the origin of this
numerical finding. First we note that the initial state at $t=0$, being the state of a system at a
first order transition point, is a mixture of clusters of the pure phases and the typical size
of the clusters, $\xi$, is finite and given by the correlation length at the first order
transition point. Now, for $t>0$, we change the form of the interaction and let relax the
system during which correlated domains of typical size $\ell(t) \sim t^{1/z}$ are created.
In the correlated volume, ${\cal V}(t) \sim \ell(t)^d$, due to random fluctuations, a given initial pure phase
has an excess volume $\Delta {\cal V}(t) \sim \ell(t)^{d/2}$ (compared to the average).
Consequently the (corrected) autocorrelation decays as $\tilde{G}(t)\sim \Delta {\cal V}(t)/{\cal V}(t)
\sim t^{-d/2z}$, so that $\lambda=d/2$, in agreement with the numerical results.

\section{Discussion}
\label{sec:disc}
In this paper we have considered a relaxation problem during which the form (and the symmetry)
of the interaction between the particles is suddenly changed, whereas the temperature of
the system is kept constant. In particular we studied such cases when the system before
and after the quench is at a critical temperature, but nevertheless belongs to different
universality classes. We have then studied well-known questions in nonequilibrium
dynamics, such as the relaxation of the magnetization and the decay of the autocorrelation
function. It is known\cite{cr97} that nonequilibrium critical dynamics at time $t$,
after a quench at $t=0$
from a state with $T_i=\infty$, is analogous to the static critical behavior of a semi-infinite
system at a distance $y$ from a free surface located at $y=0$. The analogous static
critical problem to our dynamical problem here is the interface critical behavior at a distance,
$y$, from a straight interface, which separates two coupled semi-infinite critical systems
which belong to different universality classes\cite{bti06,lti07}.
According to our numerical calculations the nonequilibrium critical behavior in our problem is the
result of the interplay and competition between the critical fluctuations of the two systems.

The specific problem we studied is the Ising model on the square lattice, with different
types of multispin interactions in the initial states, but with nearest neighbor interactions
after the quench. Two initial models, the BW model and the $n=3$ Turban model, belong to the
same static universality class. Interestingly, the magnetization scaling dimension of these
models, $x=1/8$, coincides with that of the normal Ising model. Nevertheless due to the quench,
the change of the symmetry of the interaction has a strong effect on the nonequilibrium
dynamics. Our main observation is that 
nonequilibrium dynamics has the same asymptotic behavior when we start
from initial states of different models that are in the same
(static) universality class.
For the specific problem we considered here, the nonequilibrium exponents are found to satisfy the relation
$\theta=\lambda/z$ within the error of the calculation.

The third initial model we considered, the $n=4$ Turban model, has a first-order phase transition,
thus the structure and the topology of the initial state is completely different from the
previously discussed cases. Also the
asymptotic behavior of the nonequilibrium autocorrelation function is described by a different
exponent $\lambda = 1$. 
This exponent is expected to have the value $\lambda=d/2$ and therefore to depend only on the 
dimension of the system, but to be universal otherwise, i.e. not dependent on the type of the
initial state, provided it corresponds to a first-order transition point.

\begin{acknowledgments}
This work has been supported by the Hungarian National Office of Research and
Technology under Grant No. ASEP1111, by a German-Hungarian exchange
program (DAAD-M\"OB), by the Hungarian National Research Fund under
grant No OTKA TO48721, K62588, MO45596. We are indebted to F.\'A. Bagam\'ery for cooperation
in the initial period of the project. Some part of the simulations have been done
on Virginia Tech's System X.
\end{acknowledgments}

\end{document}